\documentclass[british, 12pt, a4paper]{article}
\usepackage[left=1in,top=1in,right=1in,bottom=1in,nohead,paperwidth=8.5in, paperheight=11in]{geometry} 
\usepackage{verbatim,color,setspace,graphicx,epsfig,rotating, epstopdf,lscape, soul}
\usepackage{amsmath,amsfonts, lineno, hyperref, geometry, bbm} 
\usepackage{arydshln}
\usepackage[utf8]{inputenc}
\usepackage[british]{babel}
\usepackage{comment}
\usepackage[round]{natbib} 
\usepackage[title]{appendix}
\usepackage{ulem} 
\usepackage{rotating} 
\usepackage{amsmath,amsfonts,amssymb}
\usepackage{mathtools}
\DeclarePairedDelimiter{\ceil}{\lceil}{\rceil}

\begin{document}
\title{\textbf{Nowcasting economic activity in European regions using a mixed-frequency dynamic factor model \\
} }
\author{
Luca Barbaglia$^{a}$,
Lorenzo Frattarolo$^{b}$,
Niko Hauzenberger$^{c}$,
Dominik Hirschbühl$^{a}$,\\
Florian Huber$^{d}$,
Luca Onorante$^{a}$,
Michael Pfarrhofer$^{e}$,
Luca Tiozzo Pezzoli$^{a}$
\\
\mbox{}\\
{\small $^{a}$European Commission, Joint Research Centre}\\
{\small $^{b}$Università degli studi di Verona}\\
{\small $^{c}$University of Strathclyde}\\
{\small $^{d}$University of Salzburg}\\
{\small $^{e}$Vienna University of Economics and Business}\\
}
\date{}

\maketitle\onehalfspacing\thispagestyle{empty}\normalsize\vspace*{-2em}\small

\begin{center}
\begin{minipage}{0.8\textwidth}
\noindent\small
\textbf{Abstract}:
Timely information about the state of regional economies can be essential for planning, implementing and evaluating locally targeted economic policies. However, European regional accounts for output  are published at an annual frequency and with a two-year delay.  To obtain robust and more timely measures in a computationally efficient manner, we propose a mixed-frequency dynamic factor model that accounts for national information to produce high-frequency estimates of the regional gross  value added (GVA). We show that our model produces reliable nowcasts of GVA  in 162 regions across 12 European countries. 
\end{minipage}
\end{center}
\normalsize\renewcommand{\thepage}{\arabic{page}}

\vspace{0.4cm}
\begin{minipage}{15cm}
\singlespacing
{\footnotesize 
We thank Gary Koop, Stuart McIntyre, James Mitchell, Ping Wu, and participants of the brown bag seminar in the Department of Economics at the University of Strathclyde, as well as the participants to the JRC-DG ECFIN seminar, for insightful comments and discussions. The views expressed are purely those of the authors and should not, in any circumstances, be regarded as stating an official position of the European Commission. 
\\
\texttt{E-mail}: luca.barbaglia@ec.europa.eu (corresponding author); lorenzo.frattarolo@univr.it; niko.hauzenberger@strath.ac.uk; dominik.hirschbuehl@ec.europa.eu; florian.huber@plus.ac.at; luca.onorante@ec.europa.eu; michael.pfarrhofer@wu.ac.at; luca.tiozzo-pezzoli@ec.europa.eu.}
\end{minipage}


\newpage\doublespacing
\section{Introduction}
The timely availability of information on the state of regional economies is essential for the implementation and evaluation of targeted economic policies. 
These are often designed to address specific economic challenges or opportunities in a particular region, and their effectiveness can only be accurately assessed if policymakers have access to timely, detailed data on the economic conditions prevailing in  that region.  However,  EU-specific data collected by Eurostat  on regional output and income comes with two limitations: (1) the frequency of publication is annual, and (2) new data is released with a significant delay of two years \citep{eurostatregional}. These two shortcomings, 
combined with cross-country and cross-regional heterogeneity in terms of economic fundamentals, severely complicate monitoring and fine-tuning the impact of policy changes. 

Our model aims to address the issue of limited and delayed regional economic data by leveraging more frequently available national data.
We propose a mixed frequency dynamic factor model (MF-DFM) that uses national information, which is available on a quarterly basis, to obtain quarterly estimates of regional gross value added (GVA), which is only available at annual frequency and with a substantial time lag.  
The need to combine low and high frequency information in multivariate time series models has precedent in the literature \citep[see, e.g.,][]{mariano2003new, mariano2010coincident, schorfheide2015real, schorfheide2020real, huber2020nowcasting}.  These models are often used to obtain a monthly (or higher) frequency estimate of \textit{national} GDP, using information from higher frequency time series such as industrial production, unemployment rates or financial data. This usually involves small to medium size specifications, since standard estimation algorithms use computationally intensive filtering and smoothing steps that do not scale well to high dimensions. 

Our goal is to obtain estimates of \textit{regional} GVA at a \textit{quarterly} frequency. 
We concentrate on NUTS2-level regions, implying that the dimension of the cross-section becomes sizeable.  
This calls for a model that is capable of handling a large number of time series simultaneously. The potentially large size of the cross section is one problem, the annual frequency of our target variables is another. In our empirical work, we model quarterly movements in annual regional data 
dating back to only 2001 -- considering the required lag-structures, data availability results in $20$ observations over time for each region (in the full sample), and even shorter series when we conduct out-of-sample forecast experiments. 

In a recent paper, \citet{koop2020regional} use a model for regional GVA in the United Kingdom (UK). This model includes a moderate number of time series (the $12$ NUTS1 regions in the UK and a few additional national quantities with annual data dating back to the 1970s) and applies Markov Chain Monte Carlo (MCMC) estimation techniques to carry out estimation and inference. In addition, they introduce cross-sectional restrictions such that the sum of the regional GVA series (at the quarterly level) is equivalent to the national aggregate. Related work and refinements are provided in \citet{koop2020JRSSA,KOOP2022, koop2023incorporating}. Since our focus is on NUTS2 data for EU member states, the number of endogenous variables sharply increases which renders standard MCMC-based estimation tools impractical.\footnote{Other work at a comparable level of regional granularity often zooms into specific countries or even subsets of regions within countries, e.g., as \citet{Henzel2015,Lehmann2015,claudio2020nowcasting,lehmann2023quarterly} for Germany; \citet{Bille2023} use panel data methods for even smaller-scale considerations of Italy. A notable exception with respect to our comments about MCMC methods being infeasible is \citet{chan2023high}, who discuss computationally efficient algorithms for estimating high-dimensional state space models. A possible solution to this scalability issue is also provided in \cite{gefang2020computationally}. This paper uses variational Bayes (VB) approximation techniques to estimate mixed frequency VARs. The resulting model, however, may be hard to tune and the size of the approximation errors induced by VB-based estimation is difficult to assess.} 

In this paper, we propose a dynamic factor model that combines the virtues of the MF-DFM proposed in \cite{mariano2010coincident} with the cross-sectional and inter-temporal restrictions of \cite{koop2020regional}. We assume that the latent high frequency quantities of regional GVA co-move and feature a factor structure. This choice is motivated by the fact that regions close to each other (in geographical or economic terms, e.g., within a particular country) are often exposed to common economic shocks and these trigger substantial co-movements across regional output dynamics. A factor model is capable of handling such a behaviour in a very parsimonious manner. Since the model might be prone to overfitting if the number of factors and/or lags is large, particularly with our short time series, we use Bayesian shrinkage priors on the VAR coefficients determining the law of motion of the high frequency measures as well as the factor loadings. The resulting model can  be applied to large cross-sections since it effectively deals with computational and statistical issues by reducing the state space through factor restrictions and the introduction of highly flexible shrinkage priors.

Our resulting regional MF-DFM is then applied country-by-country to model the corresponding NUTS2 regions within a particular country.  We first provide predictive evidence that our model improves upon a simple random walk (RW) benchmark, often at sizeable margins. While we find the model to improve forecasts across all considered countries, there is some heterogeneity in predictive performance. Based on this forecast exercise, we discuss several modelling aspects that are relevant for practitioners. Finally, we investigate high-frequency dynamics of the resulting quarterly regional time series, and provide a brief (regional) case study about the onset of the COVID-19 pandemic.

The remainder of the paper is structured as follows. Section \ref{sec:econometrics} discusses the econometric framework and implementation. Section \ref{sec:results} introduces the dataset and contains our forecast exercise alongside additional empirical results. The final section concludes.

\section{Econometric framework}\label{sec:econometrics}
\subsection{The mixed-frequency dynamic factor model}
For modelling economic activity across disaggregate regions, we propose a mixed-frequency dynamic factor model (MF-DFM). Our model exploits the fact that regions within a country tend to co-move due to their exposure to common shocks.

Let $\boldsymbol{y}_t^{\ast} = (y_{1t}^{\ast},\hdots,y_{Nt}^{\ast})'$ denote a vector of observed real GVA measures for regions $i=1,\hdots,N$, where $t=1,\hdots,T$, runs on a quarterly frequency. 
This implies that $\boldsymbol{y}_t^{\ast}$ features missing values, since regional GVA is only observed annually and with a lag. We model these quantities as year-on-year differences. 
Define $\tilde{\boldsymbol{f}}_t$ as a $Q_f$-vector of latent factors, and $\boldsymbol{z}_t$ are $Q_z$-observed factors. We collect these in the $Q_f + Q_z = Q\times1$ vector $\boldsymbol{f}_t = (\tilde{\boldsymbol{f}}_t',\boldsymbol{z}_t')'$. That is, $Q \ll N$, and the framework we propose thus only relies on modelling $Q_f$ latent processes that govern within-year dynamics across regions, rather than $N$ separate processes for each individual region. While national GVA must always be included, $\boldsymbol{z}_t$ may include further information such as interest rates, unemployment, or other quantities that affect economic activity across regions.\footnote{This feature offers additional inferential possibilities such as identifying structural shocks nationally, and trace their propagation on a regional level. We leave such aspects for future research.} The variables $\tilde{\boldsymbol{f}}_t$ are never observed, while $z_{\text{GVA},t}$ are quarter-on-quarter log-differences of GVA on the national level which are observed each quarter.

We link the observed regional GVA measures to the latent and observed factors by constructing an inter-temporal restriction. For quarters when we observe annual regional GVA, we assume:
\begin{align}
y_{it}^{\ast} =& \left(\sum_{q=1}^{Q_f}\lambda_{iq}\left(\frac{1}{4}\tilde{f}_{qt} + \frac{2}{4}\tilde{f}_{qt-1} + \frac{3}{4}\tilde{f}_{qt-2} + \tilde{f}_{qt-3} + \frac{3}{4}\tilde{f}_{qt-4} + \frac{2}{4}\tilde{f}_{qt-5} + \frac{1}{4}\tilde{f}_{qt-6}\right)\right) + \nonumber\\
&\sum_{\tilde{q}={Q_f+1}}^{Q} \lambda_{i\tilde{q}} z_{\tilde{q}t} + \epsilon_{it}, \quad \epsilon_{it}\sim\mathcal{N}\left(0,\sigma_{\text{ME},i}^2\right),\label{eq:itr}
\end{align}
while in the quarters without regional observations, we have $y_{it}^{\ast} = \varnothing$. 
This equation serves two main purposes. 
First, it distributes the latent growth rates of the quarter throughout the year via the triangular weighting scheme which arises from considering log-growth rates, as proposed in \citet{mariano2003new}. Second, the unobserved and observed latent factors are then linked to observed regional growth by the factor loadings $\lambda_{iq}$ which we store in an $N\times Q$ matrix $\boldsymbol{\Lambda}$. In related papers, the inter-temporal restriction is typically specified to hold exactly; due to our factor structure, by contrast, we introduce region-specific Gaussian measurement errors centred on zero with variance $\sigma_{\text{ME},i}^2$.

A key feature of our setup is that regional growth rates can be linked to those observed quarterly on the national level \citep[see also][]{koop2020regional}. We may impose this restriction with the following measurement equation which we label the cross-sectional restriction:
\begin{equation*}
    z_{\text{GVA},t} = \sum_{q=1}^{Q_f}\left(\sum_{i=1}^N \omega_{i}\lambda_{iq} \tilde{f}_{qt}\right) + \varepsilon_t, \quad \varepsilon_t\sim\mathcal{N}\left(0,\sigma_{\text{CS}}^2\right),
\end{equation*}
where $\omega_i$ are pre-determined weights that reflect the share of the respective region in national GVA. The final aspect of our framework refers to how we model the dynamics of the observed and latent states. We achieve this by using a vector autoregressive (VAR) state equation which requires $P\geq7$ lags:
\begin{align}
    \boldsymbol{f}_t &= \boldsymbol{A}_1\boldsymbol{f}_{t-1} + \hdots + \boldsymbol{A}_P\boldsymbol{f}_{t-P} + \boldsymbol{\epsilon}_t, \quad \boldsymbol{\epsilon}_t\sim\mathcal{N}(\boldsymbol{0},\boldsymbol{\Sigma}).\label{eq:VARstateeq}
\end{align}
Here, $\boldsymbol{A}_p$ are $Q\times Q$-matrices of VAR coefficients associated with the respective $p$th lag, and $\boldsymbol{\epsilon}_t$ is a Gaussian error term with zero mean and $Q\times Q$-covariance matrix $\boldsymbol{\Sigma}$. Our measurement equations, in conjunction with this state equation, provide us with a fairly straightforward Gaussian state-space model.

\subsection{Prior setup and algorithm}
Our model is simpler than larger VAR models that explicitly feature equations for each region due to the proposed factor structure. However, the size of the cross-section and the required number of factors still yields a large number of parameters to be estimated. The latter also implies that Equation \eqref{eq:VARstateeq} may result in a comparatively large VAR. We thus regularise the parameter space via shrinkage priors, and enable equation-by-equation estimation of multivariate systems of equations.

We achieve this by estimating Equation \eqref{eq:VARstateeq} in its structural form via decomposing $\boldsymbol{\Sigma} = \boldsymbol{B}_0^{-1} \boldsymbol{H} \boldsymbol{B}_0^{-1}{'}$, such that $\boldsymbol{B}_0^{-1}\boldsymbol{\eta}_t = \boldsymbol{\epsilon}_t$ and $\boldsymbol{\eta}_t \sim \mathcal{N}(\boldsymbol{0},\boldsymbol{H})$. Here, $\boldsymbol{B}_0$ is lower triangular matrix with unit-diagonal, $\boldsymbol{H} = \text{diag}(h_1,\hdots,h_Q)$ is a diagonal matrix collecting the variances of the structural shocks.\footnote{Triangularizing the VAR this way also provides the opportunity of adding stochastic volatility (SV) with ease by introducing state equations for the elements on the diagonal of $\boldsymbol{H}$. Indeed, we have experimented with SV but found that it was unnecessary for the data used in our application.} We thus may write:
\begin{equation*}
    \boldsymbol{B}_0\boldsymbol{f}_t = \boldsymbol{B}_1\boldsymbol{f}_{t-1} + \hdots + \boldsymbol{B}_P\boldsymbol{f}_{t-P} + \boldsymbol{\eta}_t, \quad \boldsymbol{\eta}_t\sim\mathcal{N}(\boldsymbol{0},\boldsymbol{H}),
\end{equation*}
where the reduced form coefficients can be recovered from computing $\boldsymbol{A}_p = \boldsymbol{B}_0^{-1}\boldsymbol{B}_p$ for $p=1,\hdots,P$. This approach of estimating the VAR is very close to \citet{koop2020regional}, as is our prior setup which exploits the favourable shrinkage properties of a global-local prior. Specifically, we rely on regularising the $\boldsymbol{B}_p$'s, the free elements of $\boldsymbol{B}_0$ and $\boldsymbol{\Lambda}$ via Horseshoe \citep[HS, see][]{carvalho2010horseshoe} priors. Stationarity of the factors is imposed via a rejection sampling step. Standard filtering and smoothing may subsequently be carried out using the reduced form coefficients of the VAR.

The remaining priors relate to innovations in the state and measurement equations. Here, we use inverse Gamma priors on all of them. In the state equation, we have a moderately informative prior, $h_{q}\sim\mathcal{G}^{-1}(3,0.3)$ for $q=1,\hdots,Q$. On the measurement errors we impose more prior information. Since our intertemporal restriction is combined with a factor model, it is not exactly binding as in related papers. By assuming $\sigma_{\text{ME},i}^2\sim\mathcal{G}^{-1}(100,0.01)$ for $i=1,\hdots,N$, we express the belief that the restriction almost holds, although the prior allows for deviations in case it is required due to regional heterogeneity in light of the factors. Similarly, we use $\sigma_{\text{CS}}^2\sim\mathcal{G}^{-1}(5,0.05)$ on the cross-sectional restriction. This somewhat looser prior is due to measurement errors or potential revisions of the annual data.

\section{Nowcasting GVA in European Regions}\label{sec:results}

\subsection{Data overview, nowcasting design and model specification}
We study a multi-country data set, consisting of Austria (AT), Belgium (BE), Germany (DE), Greece (EL), Spain (ES), Finland (FI), France (FR), Ireland (IE), Italy (IT), Netherlands (NL), Portugal (PT) and Slovakia (SK). Following \cite{koop2020regional} our target variable is GVA and we obtain regional and national GVA figures from Eurostat regional economic accounts at \href{https://ec.europa.eu/eurostat/}{ec.europa.eu/eurostat}. We select euro area countries that have more than two NUTS2 regions and  exclude all extra-regions entities indexed by the ``ZZ'' suffix. Our final data set contains 162 NUTS2 regions in 12 European countries. Recall that GVA figures at the NUTS2 level are available on a yearly basis while they are quarterly at the country level. The annual regional figures start in 2001 and are released between February 15 and March 31 of each year, with a two-year publication delay.\footnote{For instance, the latest annual regional figures included in our exercise relate to 2021 and were published on February 20 2023. For more details, see \href{https://ec.europa.eu/eurostat/cache/metadata/en/reg_eco10_esms.htm}{ec.europa.eu/eurostat}.}

In our empirical work, we have three goals. First, we show that our proposed model produces nowcasts of yearly GVA figures that are accurate relative to a benchmark (discussed below). Second, we zoom into cross-regional and cross-country differences in nowcast performance and tease out key predictive properties of our model. Third, we discuss the qualitative features of the quarterly regional nowcasts for selected time periods (2019Q4 to 2020Q2) and across regions and countries.

To produce and evaluate our pseudo real-time nowcasts, we rely on the following design. Our first target in the hold-out sample is regional GVA in 2016. To reflect the publication schedule, we run the model using yearly data from 2001 to 2014 and quarterly data through 2016Q4. This information set is used to produce a prediction for 2016 (and a backcast for 2015 as a by-product, which we omit in our empirical discussions since our focus is on nowcasting). After obtaining the corresponding predictive densities, we add one more year of data and produce quarterly estimates of regional GVA to generate a nowcast of yearly GVA. This is done until we reach the end of our sample (annual GVA in 2021). We thus, crucially for such an exercise, respect the release calendar, but note that we do not explicitly take care of potential data revisions.

For each country in our data set, we start by estimating the MF-DFM independently and select the optimal number of factors $Q$ based on their density nowcasting performance. 
We use a grid of values for factors ranging from $1$ to the minimum between the number of regions in a country and an experimental threshold $Q_{max}$ that we set to $10$. In other words, for countries with fewer than $10$ NUTS2 regions we test the fully-specified model, while for the other countries we take 10 as the maximum number of factors. Note that the overspecified factor model closely resembles the MF-VAR model by \cite{koop2020regional}. 

In principle, modelling all countries (and associated regions) within a single model would be possible. However, we opt for not doing so for three reasons. First, while still being tractable, computation becomes much more cumbersome. This is because the space of observed series becomes much larger dimensional, and filtering algorithms turn out to be slow. Moreover, since we might have to include a more significant number of factors, the dimension of the state equation might also increase a lot. Second, model calibration becomes an issue since we introduce both cross-sectional and intertemporal restrictions. Third,  we wanted to exploit cross-country heterogeneity in the number of regions to investigate its consequences in the choice of the number of endogenous factors, something not possible in estimating a single European model.

We asses the nowcasting performance through the continuous ranked probability score \citep[CRPS, ][]{gneiting2007strictly}, weighted by the historical regional GVA shares. This provides an overall measure of regional density nowcasting performance where regions are weighted by their relative economic significance. The nowcasting results are then benchmarked against a simple random walk (RW) model specified terms of the growth rate of GVA.\footnote{As a robustness check, we also implement a standard AR($1$) model, whose nowcasting performance is comparable to the RW. The proposed MF-DFM model attains, on average, a 30\% nowcast gain with respect to both benchmarks. To simplify the exposition of the results, the main paper considers only the RW benchmark. The results for the AR model are available upon request.} 
We note that our hold-out consists of six observations, and we thus refrain from formally testing for statistical significance of differences between competing approaches.

\subsection{Nowcasting results}
We start by discussing the findings of our nowcasting exercise averaged over time. These are provided in  Table \ref{tab:qf} which reports the average CRPS by country (in columns) and number of factors $Q$ (in rows).  The figures are relative to the CRPS of the RW: values below 1 indicate a better performance of the proposed MF-DFM.  Missing values in the table indicates the countries that have fewer than 10 NUTS2 regions: for these countries, the upper threshold for $Q$ is set equal to the number of regions.

\begin{table}[]
\centering      
\begin{tabular}{c|cccccccccccc}
  \hline
    $Q$ & IE & SK & FI & PT & AT & BE & NL & EL & ES & IT & FR & DE \\ 
    \hline
    1 & 0.77 & 0.77 & 1.26 & 1.27 & 1.18 & 1.03 & 1.27 & 0.55 & 1.42 & 1.00 & 1.11 & 1.15 \\ 
    2 & 0.74 & 0.57 & 1.12 & 1.11 & 1.02 & 0.86 & 1.04 & 0.44 & 1.28 & 0.89 & 1.00 & 0.96 \\ 
    3 & 0.79 & 0.60 & 1.02 & 0.99 & 0.90 & 0.80 & 1.04 & 0.41 & 1.11 & 0.83 & 0.92 & 0.82 \\ 
    4 &  & 0.62 & 0.90 & 0.99 & 0.87 & 0.75 & 0.91 & 0.41 & 1.06 & 0.78 & 0.85 & 0.76 \\ 
    5 &  &  & 0.88 & 0.92 & 0.85 & 0.76 & 0.89 & 0.41 & 1.00 & 0.73 & 0.80 & 0.75 \\ 
    6 &  &  &  & 0.90 & 0.85 & 0.83 & 0.89 & 0.40 & 0.98 & 0.75 & 0.76 & 0.76 \\ 
    7 &  &  &  & 0.94 & 0.87 & 0.85 & 0.88 & 0.42 & 0.95 & 0.76 & 0.77 & 0.74 \\ 
    8 &  &  &  &  & 0.89 & 0.97 & 0.91 & 0.42 & 0.97 & 0.79 & 0.86 & 0.78 \\ 
    9 &  &  &  &  & 0.93 & 0.94 & 0.89 & 0.43 & 0.97 & 0.81 & 0.87 & 0.79 \\ 
   10 &  &  &  &  &  & 1.07 & 0.94 & 0.48 & 1.02 & 0.86 & 0.86 & 0.83 \\ 
   \hline
\end{tabular}

\caption{
Average CRPS across out-of-sample years (2016-2021) by country and choice of the number of factors $Q$ for the MF-DFM. 
The CRPS is weighted by the historical GVA share by country. 
The CRPS is reported relative to the one of the RW: values below 1 indicate a better performance than the benchmark.
Missing values in the table indicates the countries that have fewer than 10 NUTS2 regions (i.e., the experimental upper threshold for the grid search over the number of factors). 
}
\label{tab:qf}
\end{table}

At a general level, we observe that our model produces nowcasts that are considerably more accurate than the ones of the benchmark for most countries under consideration.  These gains vary across countries and number of factors. For some countries and values of $Q$, improvements are substantial. For instance, in the case of Greece we observe gains of approximately 60\%. In Slovakia, our model produces smaller but still substantially accuracy premia of around 40\%. For other countries (Austria, Belgium, France), we observe gains between 15 and 20\%. However, there are also some few cases where our model does not improve upon the benchmark. For instance, in Spain we find a weaker performance, with average gains below 5\% (and in fact losses for low choices of $Q$). 

When we relate the nowcasting performance \textit{within} a particular country to the number of factors, we find a consistent pattern. In case the number of factors is very low, the predictive performance of the models is poor and (with three exceptions) inferior to the one of the random walk benchmark, pointing towards underfitting. Two of these three exceptions (Ireland and Slovakia) are small countries with only a few NUTS2 regions. In this case, a single factor is sufficient to summarise the cross-sectional variation in GVA. As a general pattern, we find that the relationship between the number of factors and nowcast performance is U-shaped; the predictive performance first increases, and, if $Q$ is set too large, declines again. This points towards overfitting if the number of factors is set too large, providing some evidence that our factor model (if it is well specified) is capable of efficiently summarising the bulk of cross-regional variation. It is worth noting that the penalty in terms of predictive losses is much larger for underfitting specifications when contrasted with overfitting ones.

\begin{figure}
    \centering
    \includegraphics[scale=0.75]{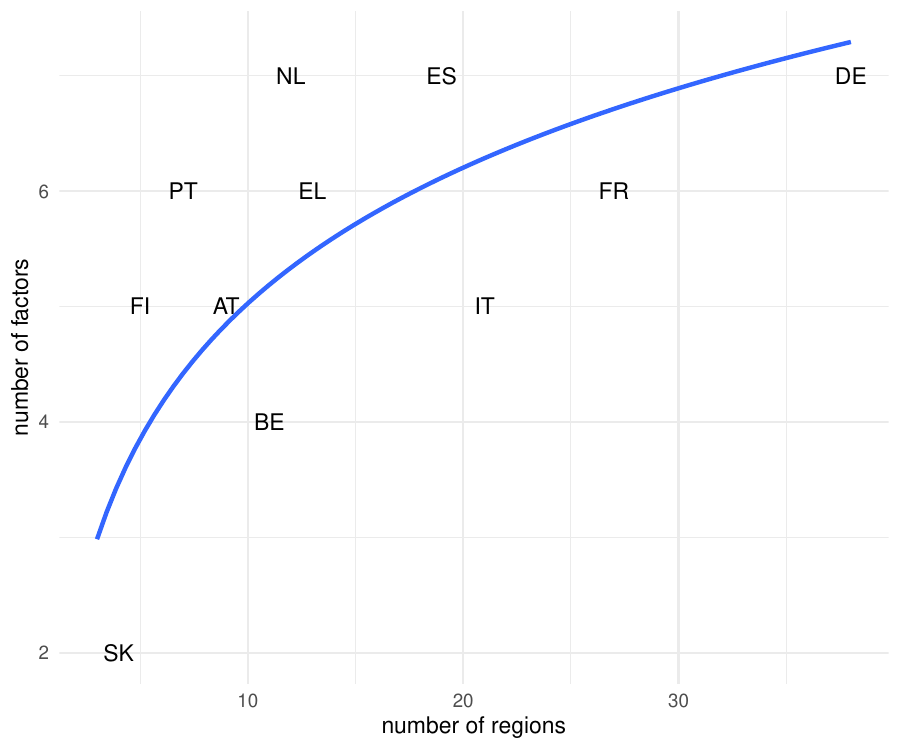}
    \caption{Optimal number of factors versus number of NUTS2 regions by country. The blue solid line is the logarithmic fit.
    }
    \label{fig:QFvsNUTS2}
\end{figure}

Our discussion also suggests that there exists a relationship between country size and the optimal number of factors. This relationship is depicted in Figure \ref{fig:QFvsNUTS2} which is a scatter plot between country size (as measured by the number of NUTS2 regions) and the optimal value of $Q$. The relationship between country size and $Q$ appears to be nonlinear and a logarithmic fit approximates this relation well. This result has important practical implications. First, few factors seems to be sufficient to summarise the information even when the nowcasting regional economic activity in large countries. Second, practitioners and forecasters can exploit the relationship between the number of regions and factors as a rule-of-thumb when setting up their nowcasting model.  A particularly simple rule that can be used in practice that exploits  this relationship sets:
\begin{equation}
 Q^* = \ceil*{\log(N)}, \label{eq: rule}
\end{equation}
where $\ceil*{\cdot}$ rounds to the next integer and $N$ is the number of NUTS2 regions within a country.
This rule thus sets the optimal number of factors as a function of $N$. In principle, this choice leads to relative CRPSs which are often extremely close to the best performing value of $Q$ in Table \ref{tab:qf}. 

\begin{sidewaysfigure}
    \centering
    \includegraphics[scale=0.68]{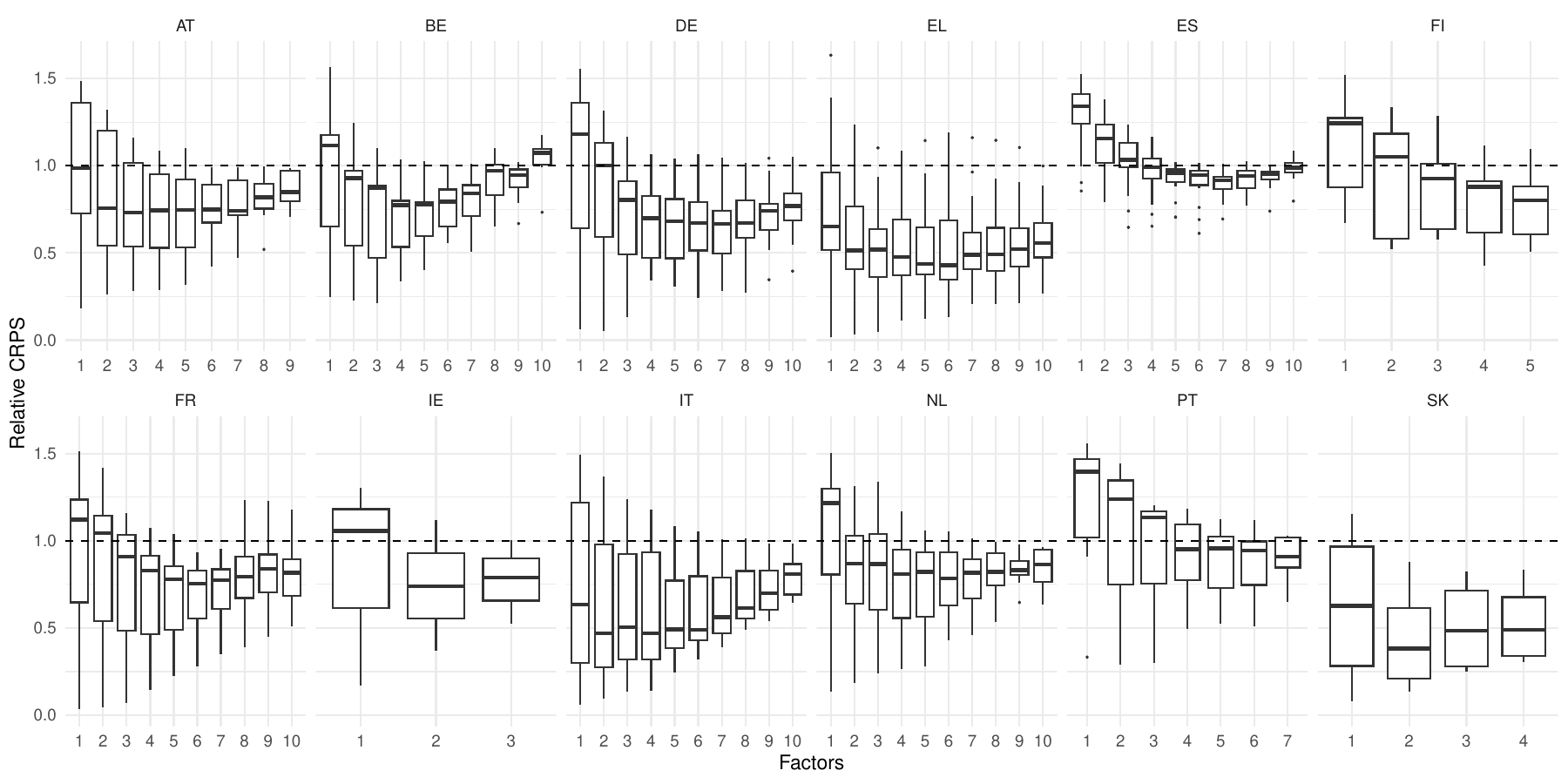}
    \caption{
     Boxplots of CRPS across number of factors for the MF-DFM models by country. 
     The maximum number of factors is limited by the number of regions in the country. 
    The CRPS is weighted by the historical GVA share across the full sample. 
    The CRPS is reported relative to the one of the RW: values below 1 indicate a better performance than the benchmark.
    }
    \label{fig:boxplot_QF}
\end{sidewaysfigure}

All results discussed so far have been based on country-specific loss measures. In the next step, we dig deeper into how nowcast performance varies with $Q$ by taking into account cross-regional variation within a given country. Figure  \ref{fig:boxplot_QF} includes boxplots for each country. The lower and upper quantiles in each box-plot represent the variation in CPRSs across NUTS2 regions and the median is the unweighted median performance per country. The figure gives rise to several relevant insights.
First, the U-shaped relationship between $Q$ and nowcast accuracy is observed once again. 
This is evidenced by the decrease in median with increasing values of $Q$ (before deteriorating slightly), as well as the downward shift in the full cross-sectional distribution of CRPSs. 
Second, the cross-sectional dispersion in forecast gains declines with increasing values of $Q$. This implies that more factors can better accommodate local dynamics, leading to a more homogeneous forecast performance for a larger value of $Q$. 
It is worth stressing, however, that this reduction in cross-sectional variance is not linear; we find a substantial reduction in variance when the number of factors is low but the effect of increasing the number of factors, if $Q$ is large, is negligible. 
Third,  adding too many factors hurts nowcast performance not only on aggregate but also for most regions under consideration.  Recall that the  fully-specified model (which sets $Q=N$ or $Q=10$) approximates  the MF-VAR by \cite{koop2020regional}, in particular if the number of NUTS2 regions is below $10$. When we compare models, we find that our proposed model achieves sizeable gains with respect to this additional ``benchmark.''  

\begin{figure}[]
    \includegraphics[scale=0.7]{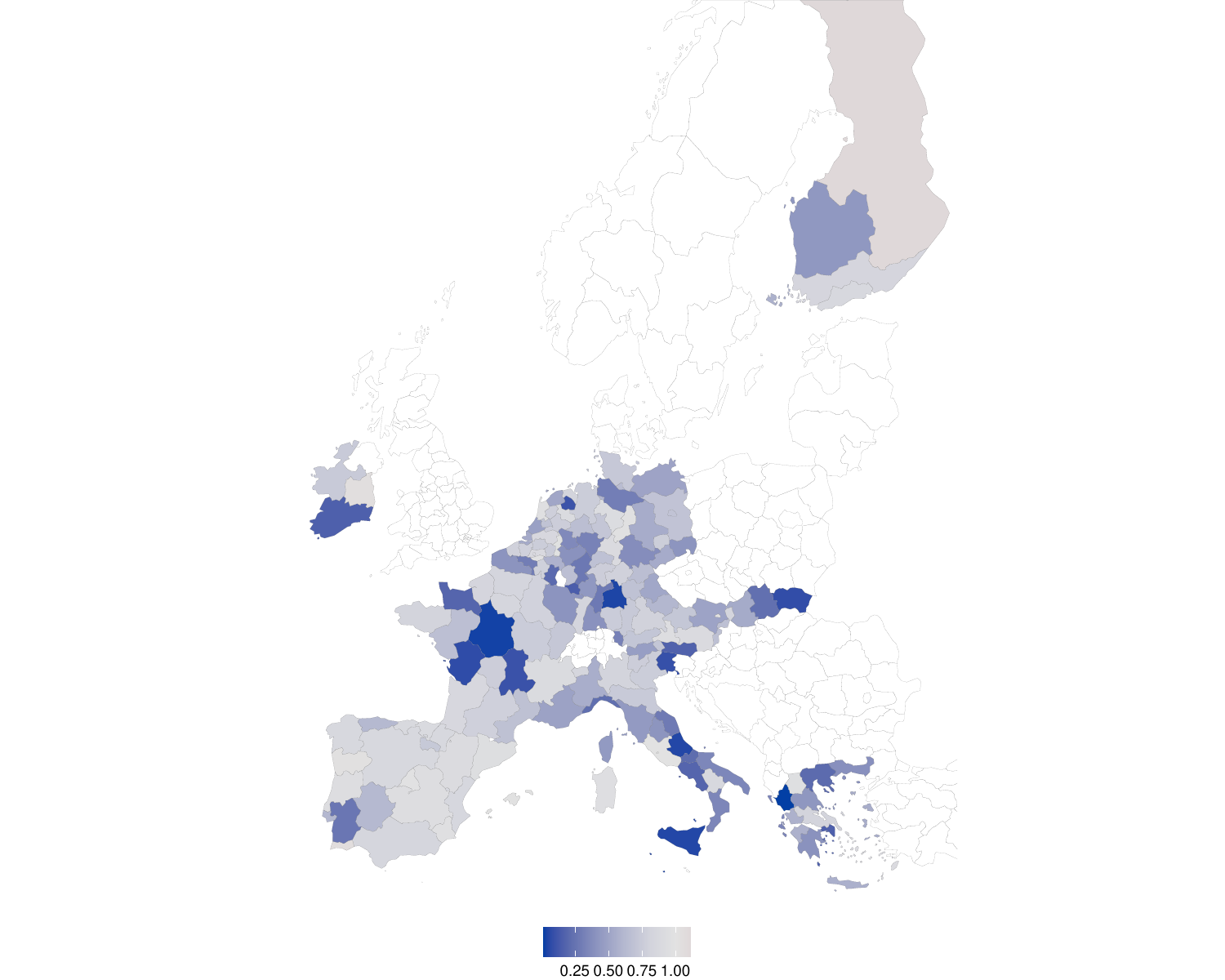}
    \caption{Average CRPS across hold-out periods (2016-2021) by NUTS2 for MF-DFM models. The CRPS is reported relative to the one of the RW: values below 1 (in blue) indicate a better performance than the benchmark.
    }
    \label{fig:map_crps_avg}
\end{figure}

The boxplots provide only a rough gauge on the overall cross-regional variation. To shed further light on how well our model (with an optimal number of factors) works in particular regions, we present a map of average nowcasting gains in Figure \ref{fig:map_crps_avg}. Regions shown in blue point towards a better performance of the proposed model with respect to the RW benchmark.

The proposed models improve upon  the benchmark in 97\% of the regions, with average gains of around 37\% across regions. This is evidenced by a great deal of blue shaded areas. Only in five out of 162 regions, the RW improves upon our proposed specification. These regions are spread across Finland, Ireland and Portugal, thus suggesting this relatively weaker performance is not country-specific. There are also several regions where improvements vis-\'{a}-vis the random walk are particularly pronounced. These are mostly located in Greece, France and Italy. Our model, with its flexibility and its use of quarterly information on national output growth, is capable of capturing this, leading to more precise nowcasts.

\begin{sidewaysfigure}[]
    \centering
    \includegraphics[scale=0.68]{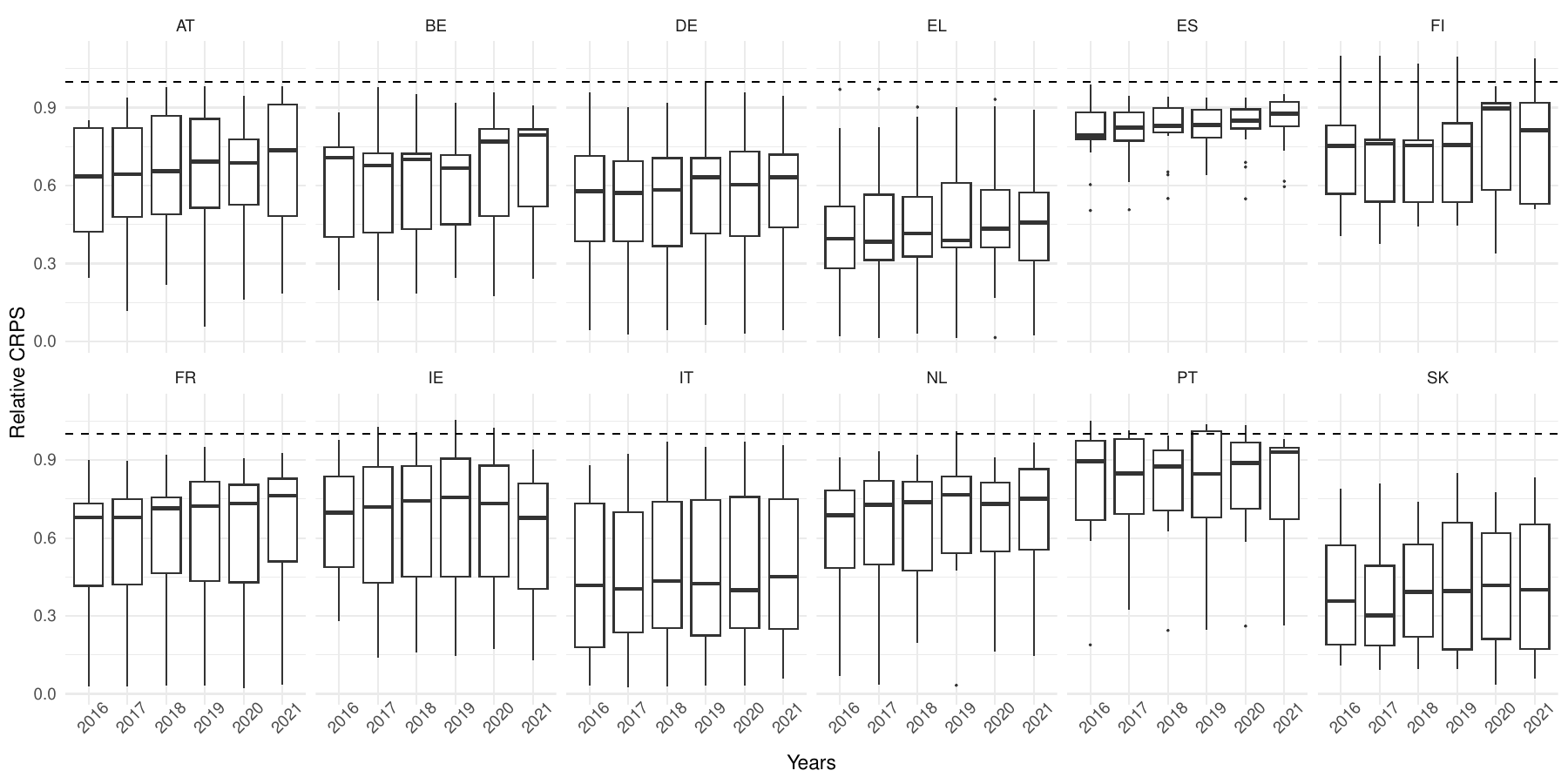}
    \caption{Boxplots of CRPS across out-of-sample years for the MF-DFM models by country. 
    The CRPS is weighted by the historical GVA share across the full sample. 
    The CRPS is reported relative to the one of the RW: values below 1 indicate a better performance than the benchmark.
    }
    \label{fig:boxplot_YEARS}
\end{sidewaysfigure}

Considering average predictive performance over the hold-out period masks information about whether our model performs differently over distinct points in time. To gain a better understanding of whether the good performance of our model is driven by a few periods in which the random walk benchmark performs poorly or whether the performance is stable and consistently better throughout the hold-out, we consider the relative CRPSs over time. These are in the form of boxplots which measure cross-sectional variation, as shown in Figure \ref{fig:boxplot_YEARS}. Again, we select the number of factors based on the average CRPS performance over time.

The figure reveals that forecast performance relative to the benchmark is remarkably stable over time. For some countries, we observe slight variation during 2020. For instance, for Austria we observe that the cross-sectional dispersion (measured through the distance between the first and third quartile) declines and substantial mass shifts downwards. A similar finding, but somewhat less pronounced, can be found for Spain. 

\subsection{Full sample estimates of quarterly regional GVA}
Important outputs of our model are quarterly series of annually released regional GVA. While the resulting time series are interesting in their own right, we also use this section to showcase some of the implications of the implementation of our model. At the heart of our framework lie the quarterly latent regional factors $\tilde{\boldsymbol{f}}_t$. These are extracted from the annually observed data using the measurement Equation \eqref{eq:itr} in conjunction with state Equation (\ref{eq:VARstateeq}). 

\begin{figure}[]
    \centering
    \includegraphics[width=\textwidth]{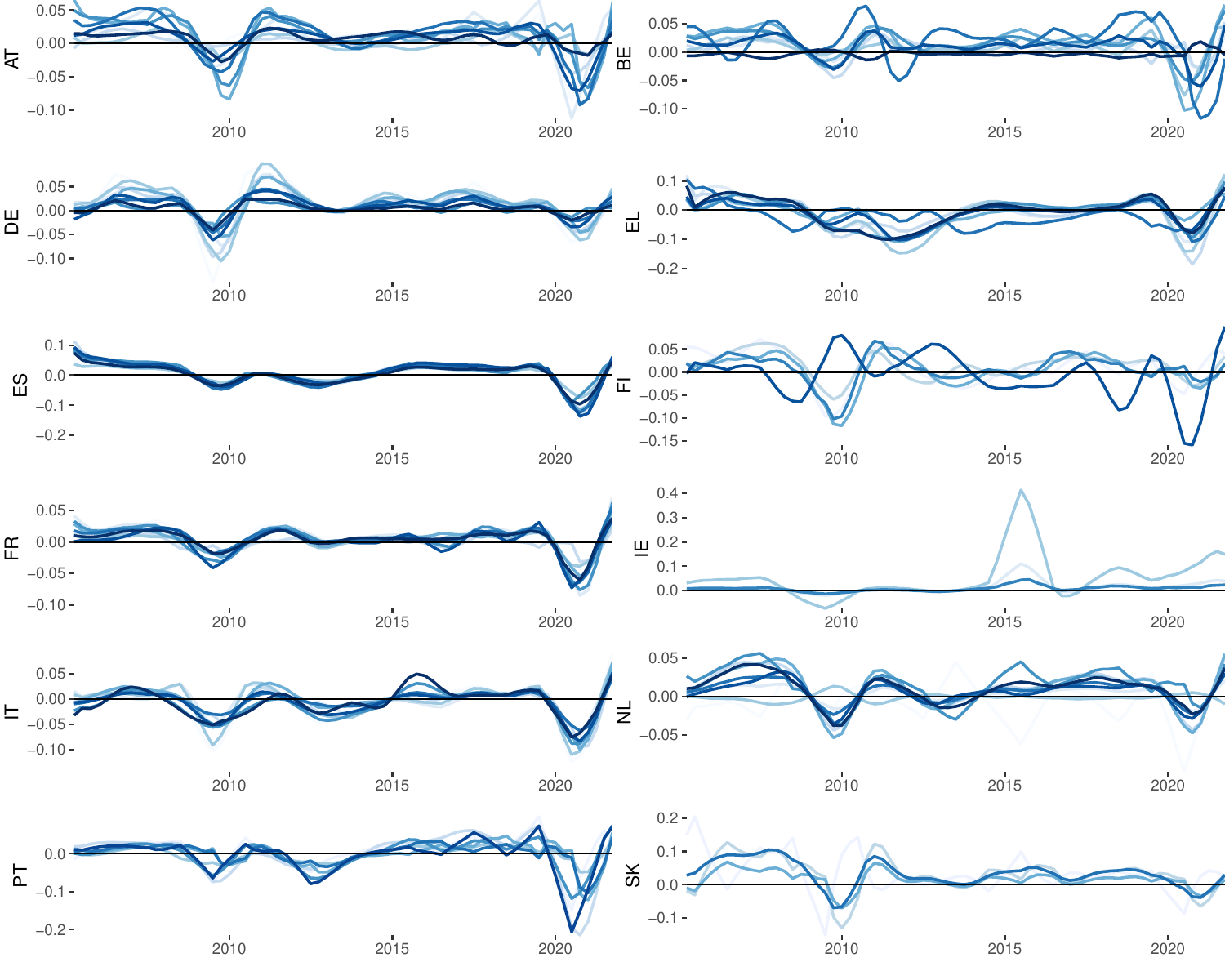}
    \caption{
    Posterior median of quarterly regional output by country (regions are shown in different shades of blue). Number of factors selected based on average forecast performance country-by-country.}
    \label{fig:regionalGVAall}
\end{figure}

The raw latent processes are not interesting by themselves, but can be transformed to the scale of the observed regional measurements via the intertemporal restriction. The ``temporal'' loadings ($1/4,2/4,3/4,\hdots$) map the latent states to annualised values, while the region-specific loadings $\lambda_{iq}$ weight these factors with respect to their importance regarding individual regions. Note that, in the absence of measurement errors, we would obtain identities for regional annual observations at the end of each year. Due to the factor structure and measurement errors we assume this is not the case, although we will see later that within-year uncertainty decreases appreciably when the intertemporal restriction holds at least approximately. The resulting posterior median estimates across all countries are shown in Figure \ref{fig:regionalGVAall}; individual regions are distinguished in different shades of blue. It is worth noting that running our model on a quarterly basis establishes a sequence of vintages of (in-sample) regional GVA series.

Overall we find that business cycle dynamics within countries (i.e., across regions for each country) are in many cases very similar, although sometimes peaks and troughs differ with respect to both timing and magnitude. Examples of such deviations are present and particularly striking in Belgium, Finland or the Netherlands. This provides evidence of the merits of our approach: substantial region-specific co-movement is modelled efficiently with the factors, but the framework is flexible enough to allow for region-specific idiosyncrasies.

Another important finding is that the quarterly regional series tend to to be correlated across countries. Since we consider independent models for the regions of each country, we do not exploit this apparent empirical regularity. We again note that extensions of our framework, e.g., featuring all European regions at once subject to country-specific and area-wide factors may be an interesting aspect for future research, but comes with several complications that we discussed earlier. A noteworthy exception to the co-movement across countries is Ireland. Irish regional output is affected by similar factors (i.e., the presence of a large number of US-multinational companies) as national GDP measurement.

\begin{figure}[]
    \centering
    \includegraphics[width=\textwidth]{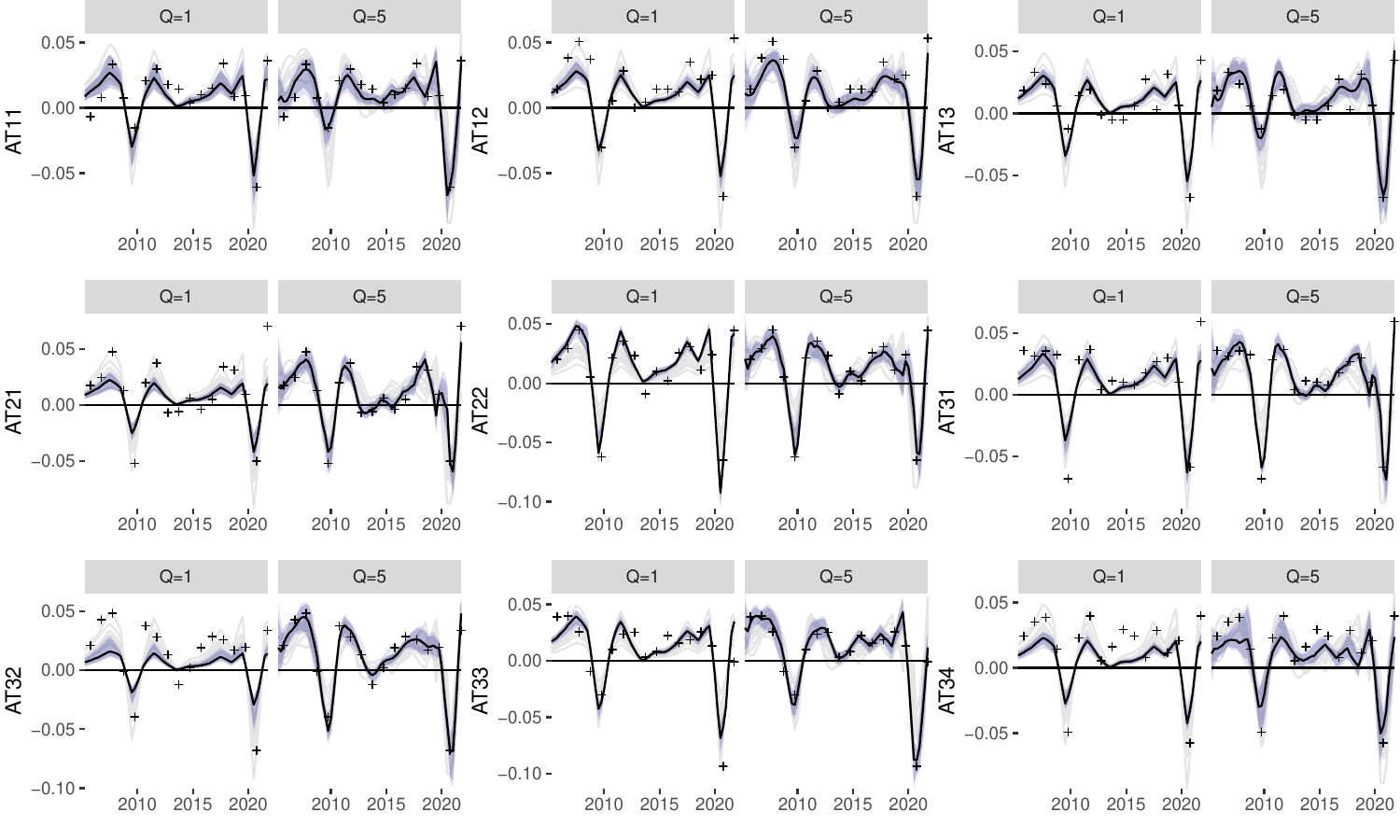}
    \caption{Regional output series for Austrian regions. Maximum number of latent factors, $Q=5$, selected based on forecast performance. Black crosses mark annual observations, shaded areas refer to 68 and 90 percent posterior credible sets. Grey lines are posterior medians for all regions other than those indicated on the y-axis.}
    \label{fig:regionalGVA-AT}
\end{figure}

To illustrate specifics about the resulting higher-frequency estimates of regional GVA, we pick Austria as an illustrative example. 
We choose this country due to its moderate number of regions and because its dynamics are representative of what we typically observe across all countries that we consider. The quarterly regional series are shown in Figure \ref{fig:regionalGVA-AT} alongside annual observations (black crosses). 
We now also indicate posterior uncertainty surrounding the point estimates of the quarterly time series. Moreover, we compare two distinct specifications of the model. First, the one-factor case, which in light of our nowcast results can be characterised as an underfitting version. Second, the multi-factor version with $Q=5$, which is selected based on predictive evidence.

Varying the number of factors has several implications for the resulting quarterly series. First, by design, the one-factor case results in the regional GVA series being proportional to one another, since they are scaled versions of the same underlying latent quarterly states. The model thus has a hard time balancing the fit across all cross-sectional units, and the actual observations in many cases are far from the estimated regional series. Second, we note that the uncertainty measures in these charts do not reflect the measurement errors, which is why the credible sets in many cases do not cover the realisations. Compared to the multi-factor model, the measurement errors are much larger in this specification as the error term compensates for the underfitting of the model. 
The shown posterior credible sets, as a consequence of $Q=1$ resulting in a small-scale model with only a moderate number of parameters, however, are tighter when compared to the $Q=5$ case. Having more factors thus increases estimation uncertainty, but also improves region-by-region fit of actual annual observations. Third, increasing the number of factors usually results in smoother estimates. This phenomenon is also visible in Figure \ref{fig:regionalGVAall} --- the resulting estimates for large countries, which have a larger number of factors, are typically smoother. Finally, as mentioned above, the intertemporal restriction is not exactly binding due to the factor structure we assume. Thus, we also observe variation surrounding estimates in quarters when regional GVA is observed.

\subsection{The onset of the pandemic through the lens of our model}
\begin{sidewaysfigure}
    \centering
    \includegraphics[scale=0.65]{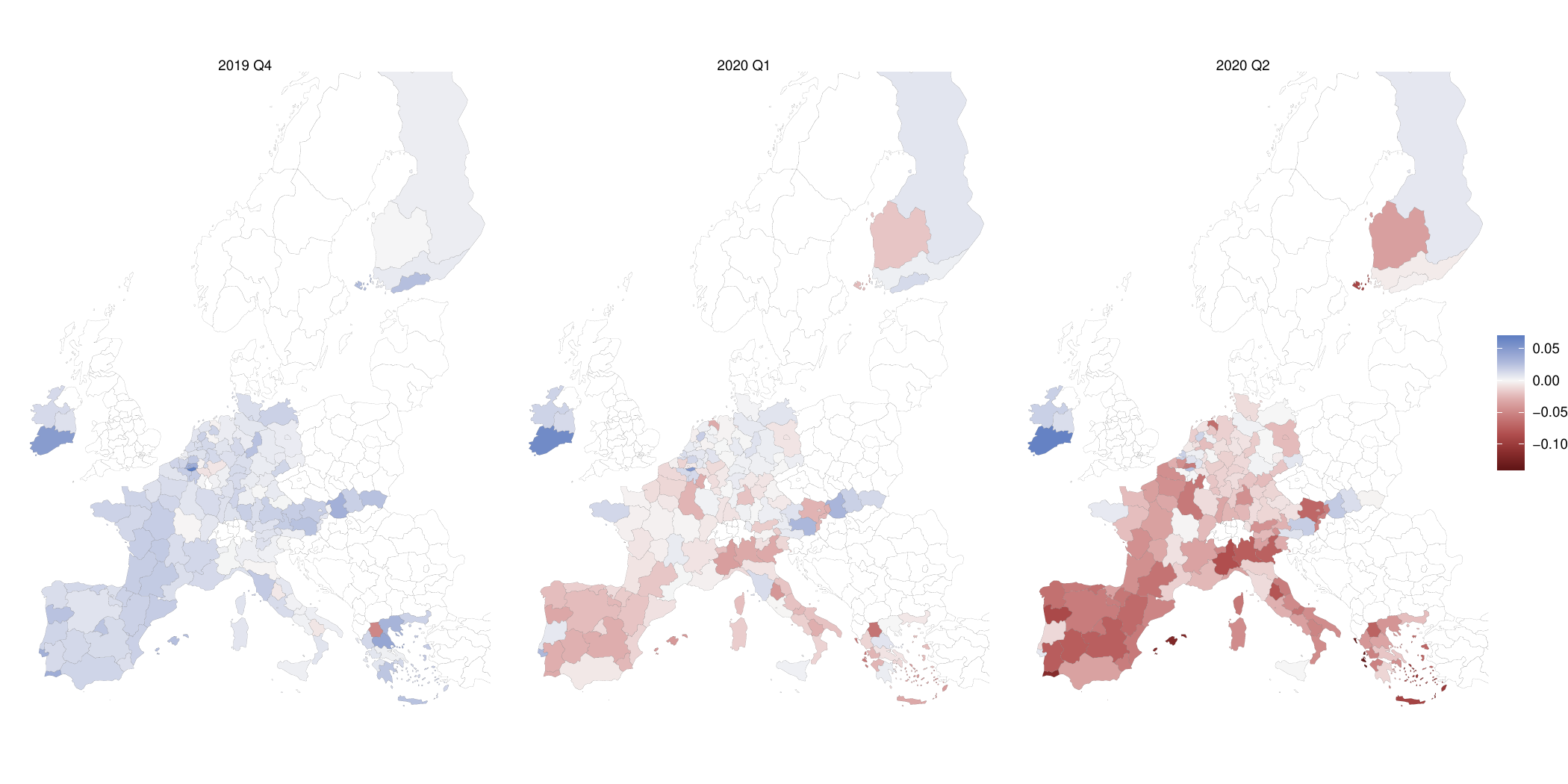}
    \caption{Quarterly estimates of regional GVA growth in 2019 Q4 and 2020 Q1-2. Negative values are in red, while positive ones are in blue.} 
    \label{fig:mapGVA2020}
\end{sidewaysfigure}

High-frequency estimates of the regional GVA can become an essential tool for policy formulation and evaluation. As an example, we investigate the behaviour of the estimated high-frequency GVA in correspondence to the breakthrough of the COVID-19 pandemic in Europe (see \citealp{carvalho2021first} for a timeline of the COVID-19 evolution). Figure \ref{fig:mapGVA2020} plots the maps of the quarterly GVA growth in the last quarter of 2019 and the first two quarters of 2020 estimated using our full sample.

The plot for 2019Q4 shows the pre-pandemic situation. While COVID-19 was already spreading at the end of 2019, its impact on economic activities materialised only starting from the first quarter of the following year. The majority of European regions experienced mildly positive output growth, except for the Greek region of Western Macedonia, the Liège region in Belgium, K{\"o}ln in Germany, and the Basilicata and Umbria regions in Italy.

The first major governmental restrictions to limit the spreading of the virus in Europe were taken at the end of February 2020 in Northern Italy and were rapidly followed by similar measures in the rest of the continent. The central panel of Figure \ref{fig:mapGVA2020} shows the estimated GVA growth in 2020Q1. Compared to the previous quarter, we observe a clear deterioration in economic conditions across Europe. The northern Italian regions of Piedmont, Lombardy and Veneto show a stronger decrease in GVA growth (the estimated GVA growth in all three regions is smaller than $-3$\%) than the rest of the sample, reflecting the fact that those regions were the first ones to be badly hit by COVID-19.

Moving to the second quarter of 2020, the situation deteriorates in all regions in our sample. Europe is now the epicentre of the COVID-19 pandemic with the highest number of reported cases and deaths worldwide, accompanied by mobility restrictions and lockdowns.\footnote{See this \href{https://www.who.int/director-general/speeches/detail/who-director-general-s-opening-remarks-at-the-mission-briefing-on-covid-19---13-march-2020}{World Health Organisation statement}.} The Northern Italian regions are again largely impacted by the negative consequences of the COVID-19 spreading, with an average estimated quarterly GVA growth of approximately $-7$\%. Other regions that show the largest contractions of GVA are some coastal regions in Portugal, as well as the insular regions in Spain and Greece: these territories are characterised by a strong dependence on the tourism and hospitality industries, which were among the most impacted by the pandemic \citep{plzakova2022impact}.

These high-frequency estimates of regional output can provide quantitative guidance for both policy formulation and evaluation in all situations where there is a need to target particular help to the most impacted regions. Moreover, in case of budgetary constraints, these estimates can help allocate the limited resources efficiently and promote the cohesion of regional policies across European territories. The data set of quarterly GVA estimates will be made available via the JRC Data Catalogue.\footnote{JRC Data Catalogue available at \href{https://data.jrc.ec.europa.eu/}{data.jrc.ec.europa.eu}.}

\section{Conclusions}
Using regional data gives rise to (at least) two challenges. First, key series such as GVA (as a measure of output) are only available on a yearly basis and released with delays of over two years. 
Second, estimating joint regional models implies that the number of time series to process becomes large. 
We address both issues by proposing a MF-DFM model. 
Our model flexibly combines national GVA data, available at the quarterly frequency, with regional GVA series that are only available at an annual frequency. We introduce temporal and cross-sectional restrictions to interpolate the yearly series and thus obtain quarterly measures of regional GVA. We thoroughly investigate the choice of the number of endogenous factors, leveraging on independent estimates for the different European countries characterised by different numbers of regions and other sources of heterogeneity. 
In a nowcasting exercise, we show that our approach yields precise density forecasts. In a case study, we show that our approach is capable of detecting the substantial decline in regional GVA during the pandemic in a timely manner.

The basic model developed in this paper can be extended along further directions. For instance, one shortcoming is that we model the countries separately. In principle, pooling all series together and thus estimating a huge dimensional factor model would be feasible. Another possible extension of our model is to allow for time-varying parameters in the state and/or the observation equation. Finally, and more from a practitioner's perspective, it is worth emphasising that our model can also be used as a tool to carry out structural inference on the regional level.

\clearpage

\small{\setstretch{0.85}
\addcontentsline{toc}{section}{References}
\bibliographystyle{jae.bst}
\bibliography{lit}}

\end{document}